\documentclass[]{sig-alternate}

\usepackage{listings}
\usepackage{graphicx}
\usepackage{booktabs}
\usepackage{todonotes}
\usepackage{hyperref}
\usepackage{algorithm2e}
\usepackage{pgfplots}
\pgfplotsset{compat = 1.9}
\usepackage{subcaption}
\usepackage{textcomp}

\newcommand*\BitAnd{\mathrel{\&}}
\newcommand*\BitOr{\mathrel{|}}

\newcommand*\BitNeg{\ensuremath{\mathord{\sim}}}

\begin{document}
%
% --- Author Metadata here ---
\conferenceinfo{CONF \textquotesingle 21}{Jan 1 -- Jan 1, 2021, CA, USA}
%\CopyrightYear{2007} % Allows default copyright year (20XX) to be over-ridden - IF NEED BE.
%\crdata{0-12345-67-8/90/01}  % Allows default copyright data (0-89791-88-6/97/05) to be over-ridden - IF NEED BE.
% --- End of Author Metadata ---

\title{Hierarchical Bitmap Indexing for Range and Membership Queries on Multidimensional Arrays
%\titlenote{(Produces the permission block, and
%copyright information). For use with
%SIG-ALTERNATE.CLS. Supported by ACM.}\titlenote{A full version of this paper is available as
%\textit{Author's Guide to Preparing ACM SIG Proceedings Using
%\LaTeX$2_\epsilon$\ and BibTeX} at
%\texttt{www.acm.org/eaddress.htm}}
}
% \subtitle{[Extended Abstract]
% \titlenote{A full version of this paper is available as
% \textit{Author's Guide to Preparing ACM SIG Proceedings Using
% \LaTeX$2_\epsilon$\ and BibTeX} at
% \texttt{www.acm.org/eaddress.htm}}}

\numberofauthors{3} %  in this sample file, there are a *total*
\author{
% 1st. author
\alignauthor
Lubo\v{s} Kr\v{c}\'{a}l\\%\titlenote{LK}\\
       \affaddr{Czech Technical University in Prague, Czech Republic}\\
       \email{lubos.krcal@fit.cvut.cz}
% 2nd. author
\alignauthor
Shen-Shyang Ho\\%\titlenote{Shen-Shyang Ho}\\
       \affaddr{Rowan University, Glassboro, NJ, USA}\\
       \email{hos@rowan.edu}
% % 3rd. author
\alignauthor
Jan Holub\\%\titlenote{LK}\\
       \affaddr{Czech Technical University in Prague, Czech Republic}\\
       \email{jan.holub@fit.cvut.cz}
% \alignauthor Lars Th{\o}rv{\"a}ld\titlenote{This author is the
% one who did all the really hard work.}\\
%        \affaddr{The Th{\o}rv{\"a}ld Group}\\
%        \affaddr{1 Th{\o}rv{\"a}ld Circle}\\
%        \affaddr{Hekla, Iceland}\\
%        \email{larst@affiliation.org}
% \and  % use '\and' if you need 'another row' of author names
% % 4th. author
% \alignauthor Lawrence P. Leipuner\\
%        \affaddr{Brookhaven Laboratories}\\
%        \affaddr{Brookhaven National Lab}\\
%        \affaddr{P.O. Box 5000}\\
%        \email{lleipuner@researchlabs.org}
% % 5th. author
% \alignauthor Sean Fogarty\\
%        \affaddr{NASA Ames Research Center}\\
%        \affaddr{Moffett Field}\\
%        \affaddr{California 94035}\\
%        \email{fogartys@amesres.org}
% % 6th. author
% \alignauthor Charles Palmer\\
%        \affaddr{Palmer Research Laboratories}\\
%        \affaddr{8600 Datapoint Drive}\\
%        \affaddr{San Antonio, Texas 78229}\\
%        \email{cpalmer@prl.com}
}
% There's nothing stopping you putting the seventh, eighth, etc.
% author on the opening page (as the 'third row') but we ask,
% for aesthetic reasons that you place these 'additional authors'
% in the \additional authors block, viz.
% \additionalauthors{Additional authors: John Smith (The Th{\o}rv{\"a}ld Group,
% email: {\texttt{jsmith@affiliation.org}}) and Julius P.~Kumquat
% (The Kumquat Consortium, email: {\texttt{jpkumquat@consortium.net}}).}
\date{30 August 2021}
% Just remember to make sure that the TOTAL number of authors
% is the number that will appear on the first page PLUS the
% number that will appear in the \additionalauthors section.

% Permission to make digital or hard copies of all or part of this work for
% personal or classroom use is granted without fee provided that copies are
% not made or distributed for profit or commercial advantage and that copies
% bear this notice and the full citation on the first page. To copy otherwise, to
% republish, to post on servers or to redistribute to lists, requires prior specific
% permission and/or a fee.
% ACM SIGSPATIAL International Workshop on Analytics for Big Geospatial
% Data 2015, Seattle, WA, USA
% Copyright 2015 ACM 978-1-4503-3974-2 ...$15.00.

% \CopyrightYear{2016}
% \crdata{------------}

\maketitle
\begin{abstract}

Traditional indexing techniques commonly employed in da\-ta\-ba\-se systems perform poorly on multidimensional array scientific data. Bitmap indices are widely used in commercial databases for processing complex queries, due to their effective use of bit-wise operations and space-efficiency. However, bitmap indices apply natively to relational or linearized datasets, which is especially notable in binned or compressed indices.

We propose a new method for multidimensional array indexing that overcomes the dimensionality-induced inefficiencies. The hierarchical indexing method is based on $n$-di\-men\-sional sparse trees for dimension partitioning, with bound number of individual, adaptively binned indices for attribute partitioning. This indexing performs well on range involving both dimensions and attributes, as it prunes the search space early, avoids reading entire index data, and does at most a single index traversal. Moreover, the indexing is easily extensible to membership queries.

The indexing method was implemented on top of a state of the art bitmap indexing library Fastbit. We show that the hierarchical bitmap index outperforms conventional bitmap indexing built on auxiliary attribute for each dimension. Furthermore, the adaptive binning significantly reduces the amount of bins and therefore memory requirements.

\end{abstract}

% \pagebreak

% \vspace{2cm}

% \category{H.4}{Information Systems Applications}{Miscellaneous}
%A category including the fourth, optional field follows...
% \vspace{-1mm}



\pagebreak
\section{Introduction}

Research in many areas, such as geoscience or model simulations, produces large scientific datasets, which are stored in multidimensional arrays of arbitrary size, dimensionality and cardinality, such as QuikSCAT satellite data~\cite{lungu2006quikscat}. Efficient processing of such data is challenging because of their multidimensional nature. However, most of the analysis techniques apply to relational datasets or require a strict linearization of the data.

To query multidimensional array data, one needs an effective system index and subsequently query the data. Majority of the current systems rely on linearization of the array data, i.e., mapping the data into one dimension, enabling many one-dimensional access methods to be used. Others, such as array databases \cite{Stonebraker2013,baumann1998multidimensional}, work natively with multidimensional arrays.

A popular and very effective method of indexing arbitrary data is bitmap indexing, which is an index consisting of a set of bitmaps (bitvectors) with associated metadata. Bitmap indices leverage hardware support for fast bit-wise operations (AND, OR, NOT, XOR), and are very space-efficient, especially for low-cardinality attributes, although this was partially overcome by sophisticated multi-level and multi-component indices. Bitmap indices are used in majority of commercial relational databases \cite{gosink2006hdf5,sinha2006bitmap,sinha2007multi,chou2011parallel}.

The major disadvantage of bitmap indices for multidimensional array data indexing is their linear nature. Even with a variation of run-length compression, of which the most well-known is WAH, that only partially suppresses the issue.

Our major contribution is a new method of bitmap indexing for multidimensional arrays that overcomes the di\-men
-siona\-li\-ty-induced inefficiencies. The method is based on $n$-di\-men\-sional sparse trees for dimension partitioning, and on attribute partitioning using adaptively binned indices. We demonstrate the performance on range queries involving both dimensions and attributes. We also show the effectiveness of our hierarchical indexing method as it prunes the search space early, avoids reading entire index data, and does at most a single index traversal.

The paper is organized as follows.
In Section \ref{sec:related_work}, we briefly describe previous work on bitmap indexing, scientific applications and multidimensional arrays.
In Section \ref{sec:preliminaries}, we describe the preliminaries to our work, including bitmap indexing, array data model and array queries.
In Section \ref{sec:hierarchical_bitmap_array_index}, we introduce our hierarchical bitmap array index, discuss its concepts, and explain its construction.
In Section \ref{sec:querying_dimensions_and_attributes}, we describe the query evaluation process for mixed attribute and dimension range queries.
In Section \ref{sec:experimental_evaluation}, we demonstrate the effectiveness on multiple queries and compare our index to other solutions.
In Section \ref{sec:conclusion}, we conclude with several notes on future research and development directions.

\section{Related Work} % (fold)
\label{sec:related_work}

Traditional indexing methods like B-trees and hashing are not effectively applicable to index multiple attributes in a single index, being replaced by multidimensional indexing methods, such as R-trees \cite{guttman1984r}, R*-trees \cite{beckmann1990r}, KD-trees, n-dimensional trees (quadtrees, octrees, etc.) \cite{samet1984quadtree,samet1990applications}. These methods are not very effective for high dimension arrays and are relatively space demanding. A good overview of spatial indexing algorithms is in \cite{samet2006foundations}, though majority of the focus is on traditional spatial data instead of multidimensional arrays.

The drawbacks of traditional indexing algorithms led to the introduction of bitmap indices \cite{chan1998bitmap} and their applications for scientific data \cite{stockinger2002bitmap}.
Bitmap indices are naturally based on linear data, ideal for relational databases. Space filling curves, such as Z-order curve and Hilbert curves \cite{lawder2000using,Lawder2001} were used for linearization and subsequent querying of multidimensional data. Hilbert curves were used in \cite{Lawder2001}, while Z-order curves were used in  \cite{Nagarkar2015}, which is a system for querying spatial data (not arrays) using compressed hierarchical bitmap indices. Hierarchically organized bitmap indices were also used for star queries on data with hierarchically organized dimensions \cite{Chmiel2010}. Bitmap indices have also been used for approximating aggregations \cite{wang2015novel}, contrast set mining \cite{zhu2015scicsm}, subgroup discovery \cite{wang2015scisd}, correlation analysis \cite{su2015situ}. All of which use bitmap indices on auxiliary attributes made from dimensions (see Section \ref{sub:bitmap_indexing}).
Other works utilize bitmap indexing for spatial applications, but do not model the data as multidimensional arrays \cite{lopes2009spatial,siqueira2012sb,stockinger2006bitmap}.

The boom of multidimensional, scientific array data gave birth to open-source multidimensional array-based data management and analytics systems, namely RasDaMan \cite{baumann1998multidimensional} and SciDB \cite{Stonebraker2013}. These databases work natively with multidimensional arrays, but lack some of the effective query processing methods implemented in other databases. On the other hand, SciDB has been established as a foundation for many multidimensional array processing tasks. Searchlight \cite{Kalinin2015} is a SciDB based system for range queries with aggregation constraints, using constraints programming on top of array synopsis -- lossy representation of small array chunks.

\section{Preliminaries} % (fold)
\label{sec:preliminaries}

We first introduce the multidimensional array data model, then describe types of commonly used queries on arrays, with some examples. Next, we introduce bitmap indexing on linear data, binning types, encoding types and compression schemes.

\subsection{Array Data Model} % (fold)
\label{ssec:array_data_model}

An array $A$ consists of \emph{cells} with \emph{dimensions} indexed by $d_1,\ldots , d_n$. Each cell is a tuple of several \emph{attributes} $a_1,\ldots,a_m$. We assume the structure of the attributes is the same for all cells in the array. The array is denoted as $A<a_1,\ldots,a_m>[d_1,\ldots,d_n]$. For example, satellite data may have latitude, longitude, altitude and time as dimensions, and precipitation, temperature, wind speed, etc. as attributes.

\begin{figure}[htbp]
    \centering
    \includegraphics[width=0.35\textwidth]{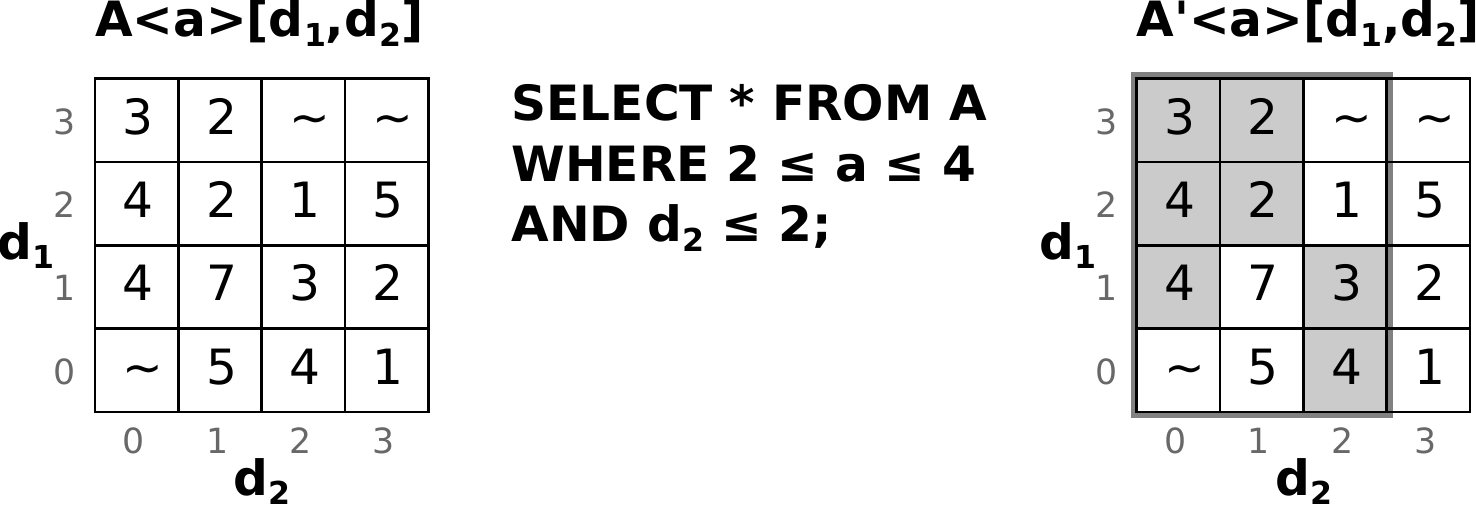}
    \vspace{1mm}
    \caption{An example of a range query on a two-dimensional array.}
    \label{fig:array-select}
\end{figure}

We form a \emph{query} on arrays based on \emph{constraints}. A dimension and attribute constraint is a constraint on a dimension and attribute in one of the following formats. A one-sided range query: $y \le 45 $; two-sided range query: $23.4 \le y < 73.2$, equality query: $y = 89$; membership query: $y \in \{2,4,6,8,10\}$, where $y$ is either dimension or attribute of the array. Figure \ref{fig:array-select} shows a query that has a two-sided constraint on an attribute $a$ and a one-sided constraint on dimension $d_2$ on a 2-dimensional array and the (shaded) query outcome. Note that equality query is a special case of membership query, and that all queries can be rewritten to a set of range queries. \emph{Mixed queries} are queries that pose constrains on at least one dimension and one attribute.

An example query on array \textsc{SatelliteArray} <snowfall, rainfall, temperature> [latitude, longitude, altitude, time] may look like this:\\
\texttt{SELECT * FROM \textsc{SatelliteArray} WHERE $50.68 \le latitude \le 50.88$ AND $14.37 \le longitude \le 14.57$ AND $30.0 \le snowfall$.}

The result would then be a possibly empty subarray of the same format as \textsc{SatelliteArray}.

\subsection{Distributed Arrays} % (fold)
\label{sub:distributed_arrays}

Due to the large size of scientific data, it is often necessary to split the data into subarrays called \emph{chunks}.

There are two commonly used strategies. Regularly gridded chunking, where all chunks are of equal shape and do not overlap. This array data model is known in SciDB as MAC (Multidimensional Array Clustering) \cite{Stonebraker2013}.
This array model works well for coarse dimension-based queries, but requires either additional indexes or filtering for fine dimension-bases and for any attribute-based queries. This array data model is the foundation (the lowest level) of our hierarchical bitmap array index. The second strategy is irregularly gridded chunking, which is one of the chunking option in RasDaMan \cite{baumann1998multidimensional}.

\subsection{Bitmap Indexing} % (fold)
\label{sub:bitmap_indexing}

Bitmap indices, originally introduced in \cite{chan1998bitmap}, were shown to be very effective for read-only or append-only data, we used in many relational databases and for scientific data management \cite{gosink2006hdf5,sinha2006bitmap,sinha2007multi,chou2011parallel}.

Bitmaps can either be created for a single attribute value, called \emph{low-level bitmaps}, or for multiple values, called \emph{high-level bitmaps}, where the bitmap is set to 1 for the cell of the arrays whose indexed value is in the value range of such bitmap.

The structure of high-level bitmaps is determined by a \emph{binning} strategy. For high cardinality attributes, binning is the essential minimum to keep the size of the index reasonable \cite{wu1998range,wu2008breaking}. Binning effectively reduces the overall number of bitmaps required to index the data, but increases the number of cells that have to be later verified. This is called a \emph{candidate check}. Two most common binning strategies are \emph{equi-width} binning, which divides the attribute domain into equal intervals, and \emph{equi-depth} binning, which divides the attribute domain into intervals covering equal (or near equal) number of cells. Equi-width binning is highly prone to excessive candidate checks, especially on skewed data.

\begin{figure}[htbp]
    \includegraphics[width=0.48\textwidth]{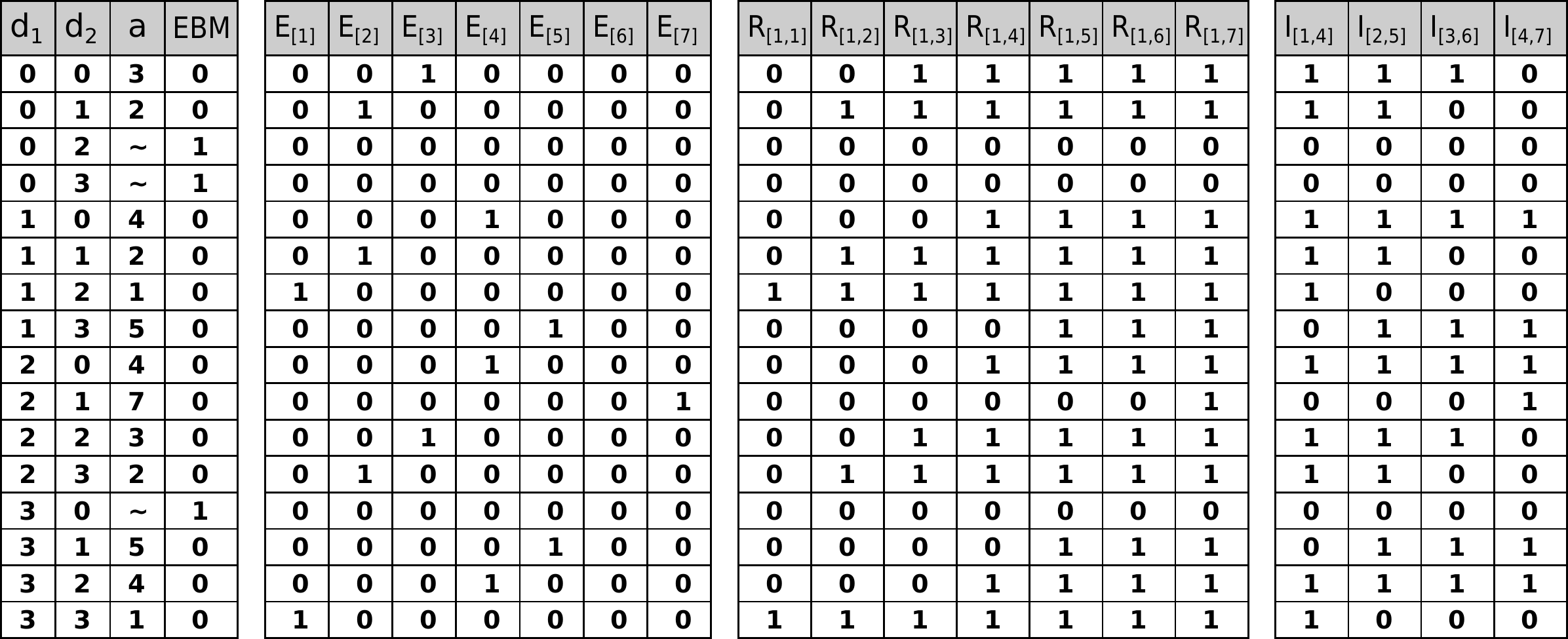}
    \vspace{2mm}
    \caption{Bitmap index for attribute \texttt{a} of the array \texttt{A} from Figure \ref{fig:array-select}: empty bitmask \texttt{EBM}, equality encoded index \texttt{E}, range encoded index \texttt{R} and interval encoded index {I}.}
    \label{fig:bitmap}
\end{figure}

Another crucial aspect of bitmap indexing is \emph{encoding} \cite{chan1998bitmap}. which determines how a set of bins, $B$, of attribute domain is encoded in each bitmap and consecutively into a bitmap index. The simplest encoding, called \emph{equality encoding}, encodes each bin with one bitmap for a total of $|B|$ bitmaps. Processing of equality queries reads a single bitmap, but processing of range queries has to read at most half of all the bitmaps. \emph{Range encoding} uses $B-1$ bitmaps, each bitmap $R_i$ encodes a range of bins $[B_1,B_i]$. The processing of range encoded bitmap index for range queries reads at most two bitmaps. \emph{Interval encoding} \cite{Chan1999} uses $\frac{|B|}{2}$ bitmaps, each bitmap $I_i$ is based on range encoded bitmaps $R_i \oplus R_{i+\frac{|B|}{2}}$. Interval encoding uses at most two bitmaps to process range queries. Compared to range encoding, it uses only half the space. Figure \ref{fig:bitmap} shows an example of equality, range and interval bitmaps for the array in Figure \ref{fig:array-select}.

Bitmap indices, based on the number of bins, may take up to $|B|\cdot C$, where $C$ is the cardinality of the indexed attribute, leading to very small number of bins needed to exceed the size of the raw data. Binary run-length compression algorithms are usually applied on bitmap indices to reduce the overall size. However, another requirement is posed to these compression algorithms, such that it must be possible to run bit-wise operations effectively on the compressed bitmaps. There are two representative compression algorithms, namely Byle-aligned Bitmap Code -- BCC \cite{antoshenkov1995byte} and Word-Aligned Hybrid (WAH) compression \cite{wu2006optimizing}.

In order to facilitate effectively high cardinality attributes with space efficient indices and fast querying, two composite methods were introduced. The first method is \emph{multi-component}, where the attribute value is decomposed into multiple components, which are then indexed independently. An example of multi-component index is a bit-sliced index \cite{o1997improved}, where each component corresponds to a bit of the value. Second composite method is called \emph{multi-level} indexing \cite{sinha2007multi}, where the binning of the attribute becomes progressively more precise with increasing levels.

Thorough performance analysis of bitmap indexing, especially multi-level and multi-component both uncompressed and compressed is presented in \cite{wu2010analyses}. An open-source bitmap indexing framework Fastbit \cite{wu2009fastbit} implements most of currently existing indexing schemes, mainly two-level indices.

\section{Hierarchical Bitmap Array Index} % (fold)
\label{sec:hierarchical_bitmap_array_index}

We now briefly discuss a common way of indexing multidimensional arrays using additional bitmap indexes for each dimension. Then we describe the structure of our hierarchical bitmap array index.

Arrays $A\langle a_1,\ldots,a_m \rangle[d_1,\ldots,d_n]$ are usually stored in a linearized representation, most commonly C-style row-major array representation. Creating one index $I_{d_i=k}(d_1,\ldots,d_n)$ for each dimension $d$, which is set to 1 for cells of array $A$ where $d$ is equal to a value $k$. This allows filtering out results based on dimensions using binary AND.

Note that the dimensions index $I_{d_i=k}(d_1,\ldots,d_n)$ does not necessarily have to use equality encoding, but based on the expected queries, we may choose a better combination of binning, encoding and compression. This approach is used in \cite{wang2015scisd,zhu2015scicsm} with equi-depth binning or in \cite{wang2015novel} with v-optimized binning based on v-optimal histograms \cite{jagadish1998optimal} and C-style row-major linearization in \cite{su2015situ}.

Unfortunately, dimension bitmap index is not effectively compressible. Consider an example of row-major ordering on 5x5 array. Then the row dimension index for $column = 1$ is \texttt{01000 01000 01000 01000 01000}, which cannot be effectively compressed using either BCC or WAH, since the compression context of both is a single bit. This can be partially mitigated by stretching dimensions to multiples of bytes or words, and extending the run-length compression to use byte or word in its compression context, instead of single bits. Another option is to use either Z-order or Hilbert space filling curves to further increase locality of the dimensions. Neither, however, solves the problem entirely.

\subsection{Partitioning of Arrays} % (fold)
\label{sub:partitioning_of_arrays}

Non-partitioned data require much finer binning and the domain of the dimension is higher than its partitioned counterpart, thus higher amount of bins is required. By partitioning the array $A\langle a_1,\ldots,a_m \rangle [d_1,\ldots,d_n]$ into a set of regularly gridded chunks $C$ in the \emph{Multidimensional Array Clustering} fashion described in Section \ref{sub:distributed_arrays}, such that:
\begin{equation*}
\begin{split}
&C_i[o_1,o_2,\ldots o_n,e_1,e_2,\ldots,e_n] =\\
&\ A\langle a_1,\ldots,a_m \rangle[o_1 \le d_1 < e_1,\ldots,o_n \le d_n < e_n]
\end{split}
\end{equation*}

All chunks in our data model are of the same shape, i.e., for all chunks $C_i, C_j$ of array $A$, it holds that
 $$C_i[e_k] - C_i[o_k] = C_j[e_k] - C_j[o_k]$$
 for all dimensions $k$, and chunks are not overlapping and completely cover the whole array $A$. In the chunk notation, $o_k$ stands for offset and $e_k$ stands for end of the chunk along that dimension (exclusive boundary).

By chunking the array, we limit the domain of both attributes and dimensions in a given partition. In our adaptive binning indices, we use the fact that the domain of the attribute varies based on the location.

The first problem arising from the equal size chunking model is that within a single chunk, we are still required to use either indexing or at least aggregate information on the attributes, such as \emph{min} and \emph{max} for precise queries or \emph{histograms} for probabilistic queries, or data exploration. We choose to use bitmap indexing on both attributes and dimensions within the chunk. Note that the dimension indices are the same for all chunks in the array, since for each chunk, we can simply subtract its offset from the dimensions query constraints.

The second problem lies in the overall structure of the chunks. There is no direct, high level index of the attributes for the chunks. It is necessary to scan through the synopsis of all the individual chunks, or generate a hierarchical synopsis. The latter has been utilized in \cite{Kalinin2015} in a form of a graph generated over merging sub-arrays.

We propose a unified solution that solves both the problem with dimension attributes and with synopsis of array chunks. Our solution is in a form of hierarchical bitmap index on top of a $n$-dimensional tree (such as octree for 3 dimensions) with variable binning for each node in the tree.

\subsection{Structure of the Array Chunk Index} % (fold)
\label{sub:structure_of_the_array_chunk_index}

The index is done separately for each attribute of the array $A$. Let's fix an attribute $\alpha$. All the following functions refer to this attribute.

Each chunk $C(o_1,o_2,\ldots,o_n)$ of array $A\langle a_1,\ldots,a_m\rangle$\linebreak$[d_1,\ldots,d_n]$ is associated with exactly one leaf\linebreak $N_\ell(o_1,o_2,\ldots,o_n)$. Independently, each leaf uses an equi-depth binning index with a total of at most \texttt{BINS} bins, where bin boundaries $bins(N_\ell$ of the index are based on an exact chunk values histogram. Note that this assumes uniform distribution of queries. If we had any prior knowledge of the queries based on the attribute, we would instead opt for weighted histogram to construct the binning.
The leaf's dimension boundaries correspond to its associated chunk's boundaries, clipped by the global shape of the array $A$.

Accounting for empty values (missing cells in $A$) is done using a special bitmask, known as \emph{empty bitmask}, for a total of $\mathtt{BINS}+1$ indices. Only leaves with at least $E \cdot \mathtt{BINS}$ non-empty cells are indexed, where the constant $E$ is dependent on the data structure used for the leaf representation, i.e., do not use bitmap indexing if listing the values is more space efficient.

Encoding of the leaf indices is left as a parameter to the user, as the bitmap indexing performance heavily depends on the cardinality of the array attribute, desired number of bins, and query types. For generality, we assume high cardinality attributes, such as integers and doubles and small number of bins such as $\mathtt{BINS} \le 16$.

Except for very narrow dimension range queries, a dimension query will either cover the whole span of a leaf node, or result in a one-sided dimension range query once the query processing reaches a single chunk. Thus, the ideal encodings for chunks are \emph{range} and \emph{interval} encodings \cite{Chan1999}. Our default encoding is  interval encoding since it uses half the memory range encoding does. Encoding of inner nodes is more complicated and we describe it in Section \ref{sub:encoding_of_bitmap_indices_in_internal_nodes}.

\subsection{Structure and Construction of the Hierarchical Bitmap Array Index} % (fold)
\label{sub:structure_and_of_the_hierarchical_bitmap_array_index}

To deal with the higher level index, we create a special composite index on tree similar to $n$-dimensional tree. Each internal node of the index has at most $F$ children, where $F$ is called a \emph{fanout}. Note that, unlike in quadtrees, octrees or $n$-dimensional trees, $F$ is not necessarily $2^n$, where $n$ is the number of dimensions. Our bitmap indices are based on the fanout and we want to utilize binary operations as much as possible. For this reason, the fanout $F$ should be a multiple of the processor word size $W$, or as close to it as possible.

The overall internal node fanout $F$ can be expressed in terms of a fanout $F_{d_k}$ for a single dimension $k$ as
$$F = \prod_{k=1}^{n} F_{d_k} \le \left( \max_{1 \leq k \leq n} F_{d_k} \right)^n$$

Assuming that the dimension fanout $F_{d_k}$ is the same for all dimensions, we can get
$$F_{d_k} = \left\lfloor F^{\frac{1}{n}} \right\rfloor$$

As we will see in Section \ref{sub:dimension_based_matches}, in order to facilitate efficient dimension range queries, the size of $F$ cannot be too large, since the size of precomputed dimension clipping bitmaps depends on $F$.

The index tree construction works in a bottom-up fashion, where the leaf nodes are indexed at first. This allows both data appending and modification (see Section \ref{sub:appending_and_modifying_data}). Each internal node is constructed from at most $F$ direct children and with at most $\mathtt{BINS}$ attribute bins, with one additional index for empty bitmask. Each child node $N_i$ of internal node $N$ provides its attribute's $min(N_i)$ and $max(N_i)$ values. These values are used for the construction of the bitmap index of $N$.

Let $B = {(min(N_1),max(N_1)),\ldots,(min(N_F),max(N_F))}$ be the set of all intervals ranging from the minimum to the maximum value of the indexed attribute $\alpha$ among all the child nodes $N_i$. The set $B$ is the set of bins -- the individual interval boundaries are delimiters, where the attribute's $\alpha$ value $a$ is in the attribute domain of different child nodes. Formally, let $nodesin(a) \subset {N_i}$ be a function of a value $a \in \alpha$ of attribute $\alpha$, which returns a subset of child nodes.
$$ N_i \in nodesin(a) \iff min(N_i) \le a \le max(N_i) $$

The set $nodesin(a)$ is used to construct the binning for index of this internal node. We describe the encoding of this bitmap index in Section \ref{sub:encoding_of_bitmap_indices_in_internal_nodes}.

The index bins are aligned with the bins from $B$. This guarantees that no two indices for different bins will be identical, i.e., represent the same set of children. It also directly implies that adding more boundaries to $B$ would be pointless.

\subsection{Bin Boundaries Merging in Parent Nodes} % (fold)
\label{sub:bin_boundaries_merging_in_the_parent_node}

\begin{figure}[htbp]
    \centering
    \vspace{2mm}
    \includegraphics[width=0.48\textwidth]{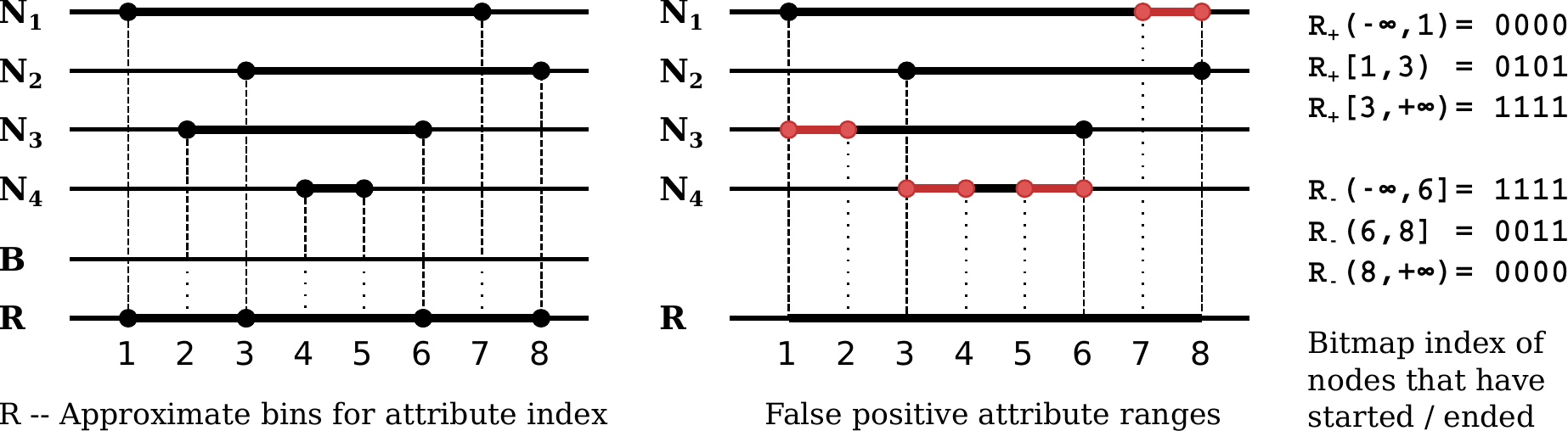}
    \vspace{2mm}
    \caption{Example of merging $|B|=8$ bin boundaries to $|R|=4$ bin boundaries for 4 child nodes. False positive ranges are marked in red. Two sided range encoded bitmaps are generated for $R$.}
    \label{fig:ranges}
\end{figure}

The number of bins from all $F$ child nodes is higher than \texttt{BINS} for majority of the internal nodes $N$, therefore it is necessary to reduce the size of the set of bins, $B$. There are several strategies to choose $B \subset D$ such that $|B| = \mathtt{BINS}$. An example of such binning reduction is in Figure \ref{fig:ranges}.

The first strategy is to use an equi-width distribution of the bins. This is the ideal choice assuming the attribute part of the query is uniformly distributed or when there is no prior knowledge about the attribute query and assuming the data distribution is not skewed.

The second strategy is to use equi-depth binning. This is ideal if the attribute distribution of the child nodes is skewed. It is possible to maintain the weights of the bins for leaf nodes, since those have direct access to the data. However, internal nodes can only make estimates about the weight of merged bins. In each internal node and leaf, we store the weight estimate $w(b)$, where $b \in B$. The weighted square error of a bin $b$ is
$$ wse(b) = \left|w(b) - \frac{w(D)}{\mathtt{BINS}}\right|^2 $$
and the weighted sum square error is
$$wsse(B) = \sum_{b \in B} wse(b)$$.

To estimate the weight of merged bin $r \in R \subset B$, we assume uniform distribution of values over the intervals of bins $b \in B$. Then the estimated weight of $r$ is
$$ w(r) = \sum_{b \in B}{w(b)\cdot\mathtt{sizeof}(b \cap r)} $$
where \texttt{sizeof}$(b \cap r)$ is the size of the intersection of $r$ and $b$.

We cannot use the trivial algorithm for equi-depth binning, because we can only iterate by bins of variable weight, instead of iterating by single data points. This is why we need to approximate the equi-depth using a simple iterative algorithm. Details on selecting $R \subset B$ approximately equi-depth bins are shown in Algorithm \ref{algo:iterative_equi_depth}. We first start with equi-width binning (line 1). Then, we  generate sets of all possible bin splits and merges (lines 2-3), setup two priority queues and evaluate all possible splits and merges in terms of weighted sum square error (lines 4-11). After that, we perform one valid split and one merge on the binning as long as this leads to an improvement of the overall binning (lines 14-18). This preserves the total number of bins.

\begin{algorithm}
\LinesNumbered
 \KwIn{set of bins $B$, set of weights $w(b)$, $b\in B$, number of output bins $\mathtt{BINS}$}
 \KwResult{approx equi-depth bins $R \subset B$, $|R| = $\textit{BINS}}
 \BlankLine
 $R \leftarrow $ eq-width bins from $B$, $|B|=$\textit{BINS}\;
 $B_S \leftarrow $ all possible split bins of $R$\;
 $B_M \leftarrow $ all possible merged bins of $R$\;
 $Q_{SPLIT} \leftarrow $ priority\_queue()\;
 $Q_{MERGE} \leftarrow $ priority\_queue()\;
 \For(\tcp*[h]{bins to split}){$s \in B_S$}{%
    add $(s, \Delta wse(s))$ to $Q_{SPLIT}$\;
 }
 \For(\tcp*[h]{bins to merge}){$(m, m') \in BM$}{%
    add $((m,m'), \Delta wse((m,m'))$ to $Q_{MERGE}$\;
 }
 \tcp{split that decreases wsse the most}
 $(s, \Delta wse(s) \leftarrow min(Q_{SPLIT})$\;
 \tcp{merge that increases wsse the least}
 $((m,m'), \Delta wse((m,m'))) \leftarrow min(Q_{MERGE})$\;
 \While{$\Delta wse((m,m')) > \Delta wse(b)$}{
  split $b$\;
  merge $(b,b')$\;
  update $R, B_S, B_M, Q_{MERGE}, Q_{SPLIT}$\;
 }
  \vspace{1mm}
 \caption{Iterative equi-depth binning approximation}
 \label{algo:iterative_equi_depth}
\end{algorithm}

In case a node has either a low cardinality attribute throu\-ghout all the child nodes, we create bins mapped to single values of the attribute and their corresponding bitmaps.

Note that v-optimal binning does not work in our case, since we don't have the individual data values available during construction of the internal nodes, although we could approximate this using uniformly or normally distributed estimates within the bins of child nodes, or by propagating at least basic data synopsis.

\subsection{Double Range Encoding of Bitmap Indices in Internal Nodes} % (fold)
\label{sub:encoding_of_bitmap_indices_in_internal_nodes}

Unlike in bitmap indexing in leaves where one encodes positions of individual values, we encode sets of child nodes $nodesin(a)$ for attribute values $a$ in the internal nodes. Our binning $B$ has the property that for all attribute values $a_b,a_{b}'' \in b \in B$ it holds that $nodesin(a_b) = nodesin(a_{b}')$. Note that this does not hold for intervals $r \in R$ (See Figure \ref{fig:ranges} for an example).

We will now describe an effective bitmap encoding of\linebreak $nodesin(a)$, $a \in r \in R$. Let's have two adjacent intervals $r \in R$ and $r' \in R$, such that $r_h = r_\ell'$ Note that since $R \subset B$, we have $nodesin(r) \ne nodesin(r')$.
If $nodesin(r') \supset nodesin(r)$, then $r'$ corresponds to a bin, where nodes are added, and we add $r'$ to a set $R_+$. Else, if $nodesin(r') \subset nodesin(r)$, then nodes are removed in set $nodesin(r')$, and we add $r'$ to set $R_-$. Otherwise, some nodes are added and some are removed and we add $r'$ to both $R_+$ and $R_-$. In our example in Figure \ref{fig:ranges}, $R_+ = \{[1,3),[3,6)\}$ and $R_-=\{(3,6],(6,8]\}$.

There is no guarantee that $|R_+| = |R_-|$. If we wanted, we could run Algorithm \ref{algo:iterative_equi_depth} separately on boundaries $B_+$ and $B_-$ (likewise defined) and with $\frac{\mathtt{BINS}}{2}$ bins, but then we'd lose the equi-width approximation.

Now, we encode $|R_+|+1$ bitmaps using range encoding, so that the index for bin $r_+ \in R_+$ corresponds to children, whose attribute range minimum $min(N_i)$ is $\le$ to the upper boundary of interval $r_+$. In our example, bitmap corresponding to $r=[1,3) \in R_+$ is \texttt{0101}, indicating that $N_1$ and $N_3$ have started in or before this interval. Similarly, we encode $|R_-|+1$ bitmaps for values $r_-$ using inverse range encoding, i.e., children, whose attribute range maximum $max(N_i)$ is $>$ to $r_-$ are encoded by 0 in the bitmap, representing children that have already ended before or in the interval $r_-$.

These two bitmaps easily allow evaluation of partial and complete matches (see Section \ref{sub:attribute_based_matches}) using only two bitmap reads and one logical operation for both partial and complete query.

\subsection{Locality of the Hierarchical Index} % (fold)
\label{sub:locality_of_the_hierarchical_index}

In order to preserve locality of the data during queries, we store the whole index in a locality preserving linearization of an $n$-dimensional tree. For each query, blocks of the index are loaded sequentially and sparsely, based on the parameters in the query. Thus, only one traversal, possibly incomplete, of the index data is needed. The index data consist of bin boundaries, weight estimates and bitmap indices.

We use space filling curves, namely the Z-order curve to linearize the multidimensional array index. We choose not to use recursive multi-level Z-order curves, as this would force the query processing to be based on pre-order traversal of the index tree. We also choose not to use row major ordering, since it has poor locality and it would slow down retrieving locations child nodes and partitions. Hilbert curve has perfect locality, but it does not preserve dimensions ordering. This means we would need to precompute bitmaps for dimension constraints for each block of Hilbert curve separately. Z-order curve allows for fast child and parent node index computations, preserves dimensionality between different level and has a good locality.

The order $\mathcal{Z}_\ell$ of the Z-order curve of level $\ell$ is determined by the maximal fanout $F_{max} = \max_{1 \leq k \leq n} F_{d_k}$, where $F_{d_k}$ is a fanout of dimension $k$.
$$ \mathcal{Z}_\ell = \ell\cdot\left\lceil \log_2{F_{max}}\right\rceil $$
Assuming $F_{d_k}$ is the same for all dimensions, the order of Z-order curve is then
$$ \mathcal{Z}_\ell = \ell\cdot\left\lceil\log_2{\left\lfloor F^{\frac{1}{n}}\right\rfloor}\right\rceil $$
and such a Z-order curve has length of $(\mathcal{Z}_\ell)^n$.

Several of the higher levels are stored in a dense vector, as specified by a user parameter. These vectors are expected to be densely filled. The remaining levels are stored as non-overlapping intervals on a Z-order dimension (1D) in continuous blocks, indexed by a binary search tree. This is a compromise between sparse single node map and full vector used for higher levels. Note that the blocks may not be sequential in memory, but at most a single transition is guaranteed, i.e., no blocks are read twice during the processing of a single query.

\subsection{Appending and Modifying Data} % (fold)
\label{sub:appending_and_modifying_data}

Scientific data is often considered either fixed or append only, our indexing approach allows for both appending and data modification, although the latter is not convenient.

To append data along any dimension, we apply the same bottom-up procedure to update the index. It is necessary to update the dimension bounds of internal nodes (that were possibly previously clipped by the global shape of the array) and bitmap indices (to include the new child nodes). Note that we do not have to update the weight estimates and bin boundaries (except min and max) in order to assure index correctness. However, in order to assure the equi-depth optimal binning, we need to run the bin merge algorithm again on affected nodes.

\section{Querying Dimensions And\\ Attributes} % (fold)
\label{sec:querying_dimensions_and_attributes}

In this work, we focus on selection queries over dimensions and attributes of an array. Such query consists of a set of dimension constraints and attribute constraints. Let's specify a query $q$ over an array $\mathcal{A}\langle a_1,\ldots,a_m \rangle [d_1,\ldots,d_n]$ as a set of ranges over dimensions $q_{D}$ and attributes $q_{A}$.
$$ q = q_{A} \cup q_{D} = \{(a,a_\ell,a_h)\} \cup \{(d_j,j_\ell,j_h),\ldots\} $$
where $(a,a_\ell,a_h)$ is a triple specifying attribute constraint: attribute, its lower bound and its (exclusive) upper bound; same goes for dimensions. In this work, we focus on a single attribute query. Therefore, we simplify $q_A$ to $(a_\ell,a_h)$. It is possible for a query to not specify constraints for some dimensions, in which case we fill all $q$ with remaining dimensions, to a complete query. Dimensions, that were not specified, are filled with $(d_j,min(d_j),max(d_j))$ triples. One-sided range constraints are also extended in similar manner.

The core of the query algorithm is a breadth-first descent through the index tree. At each level, the search space is pruned according to both dimension and attribute values.

Let $N$ be the currently searched node, $N_i$ be its child nodes, where $0 \le i < F$; multidimensional range $D_N$ be the set of dimension boundaries in the format $[D_N[d]_\ell,D_N[d]_h]$, where $d$ is dimension, $\ell$ designates lower bound, $h$ upper bound, associated with node $N$.

Throughout the query processing, we maintain a queue of partially matched nodes $P$ and a set of completely matched nodes $C$. We start at a root node $N_r$, setting $P = \{N_r\}$, assuming that both: node $N$'s boundaries and query dimensions are not disjoint: $D_N \cap Q_D \ne \emptyset$ and \linebreak$(min(N),max(N)) \cap Q_A \ne \emptyset$, otherwise node $N \notin P$ and $N \notin C$.

Let $p$, $p'$, $p*$ and $c$, $c'$, $c*$ be zero bitmaps of size $F$; the bitmaps $p$ indicates partial attribute matches among the children of node $N$, $p'$ indicated partial dimensions matches, $p*$ indicates partial matches, similarly the vectors $c$, $c'$, $c*$ indicate complete matches. We will now set these vectors according to the query $Q$ for the first node in queue $P$. The partial and complete matches bitmap computation is also described in Algorithm \ref{algo_query} and in Figure \ref{fig:partial-complete-query}.

\begin{algorithm}
\LinesNumbered
 \KwIn{
    query $q = \{(a_\ell,a_h),(d_1,d_\ell,d_h),\ldots\}$ with \texttt{DIMS}\\ dimension constraints;
    node $N$;
    node children $N_1,\ldots,N_F$;
    boundaries $[D_N[d]_\ell,D_N[n]_h]$ for $N$ and all $N_i$ and dimensions $d$;
    }
 \KwResult{
    partial matches $p*$;
    complete matches $c*$;
 }
 \BlankLine

 $\mathcal{P}_{N,\mathcal{S}},\ \mathcal{C}_{N,\mathcal{S}} \leftarrow $ load index for node $N$\;
 $\mathcal{P}'_{\mathcal{S},d},\ \mathcal{C}'_{\mathcal{S},d}$;\ \ \tcp*[h]{precomputed}\;

 $p \leftarrow \{0\}^F, p' \leftarrow \{0\}^F, p*$\;
 $c \leftarrow \{1\}^F, c' \leftarrow \{1\}^F, c*$\;

 \uIf{$a_h < min(N)$ or $a_\ell > max(N)$}{
    \KwRet{$p* \leftarrow \{0\}^F$, $c* \leftarrow \{0\}^F$}
 }

 $c = c \BitAnd \mathcal{C}_{N,\mathcal{S}}(a_\ell,a_h)$\;
 $p = p \BitOr \mathcal{P}_{N,\mathcal{S}}(a_\ell,a_h) \BitAnd \BitNeg c$\;

 \For{dimensions $d$, $1 \le d \le \mathtt{DIMS}$}{
     \uIf{$d_h < D_{N_i}[d]_\ell$ or $a_\ell > D_{N_i}[d]_h$}{
        \KwRet{$p* \leftarrow \{0\}^F$, $c* \leftarrow \{0\}^F$}
     }
     \uIf{$d_\ell > D_N[d]_\ell $}{
        $p' = p' \BitOr \mathcal{P}'_{\mathcal{S},d}(d_\ell)$\;
     }
     \uIf{$d_h < D_N[d]_h $}{
        $p' = p' \BitOr \mathcal{P}'_{\mathcal{S},d}(d_h)$\;
     }
     $c' = c' \BitAnd \mathcal{C}'_{\mathcal{S},d}(d_\ell,d_h)$\;
 }
 $p' \leftarrow p' \BitAnd c'$\;
 $c' \leftarrow c' \BitAnd \BitNeg p'$\;
 $c* \leftarrow c \BitAnd c'$\;
 $p* \leftarrow (p \BitOr c) \BitAnd (p' \BitOr c') \BitAnd \BitNeg c*$\;
 \KwRet{$p*$, $c*$}
 \vspace{1mm}
 \caption{Evaluation of partial and complete match bitmaps for a single node.}
 \label{algo_query}
\end{algorithm}

\subsection{Attribute based Matches} % (fold)
\label{sub:attribute_based_matches}

In this subsection, we explain how attribute bitmask is set. This subsection further describes lines 5--8 in Algorithm \ref{algo_query}.

If $a_h < min(N)$, or $a_\ell > max(N)$, there are neither partial nor complete attribute matches and we terminate processing the current node.

Let $\mathcal{P}_{N,\mathcal{S}}(a_\ell,a_h)$ be a \emph{partial attribute match} bitmasks specific to node $N$ of for an array of shape $\mathcal{S}$, with bits set to one corresponding to children $N_i$ so that the intersection $[a_\ell,a_h] \cap [min(N_i),max(N_i)] \ne \emptyset$.
\begin{align*}
\mathcal{P}_{N,\mathcal{S}}(a_\ell,a_h)[i] &= 1 \iff \mathcal{P}_{B|N,\mathcal{S}}(a_h)[i] \land \neg \mathcal{P}_{E|N,\mathcal{S}}(a_\ell)[i] \\
\mathcal{P}_{B|N,\mathcal{S}}(a)[i] &= 1 \iff min(N_i) \le a\\
\mathcal{P}_{E|N,\mathcal{S}}(a)[i] &= 1 \iff max(N_i) \ge a
\end{align*}
The second expression describes bitmap set to 1 for children that have started before or at value $a$, the third one describes children that have ended at or after $a$. The first expression then combines both.

To evaluate $\mathcal{P}_{N,\mathcal{S}}(a_\ell,a_h)$, we first use binary search on $R_+$ and $R_-$ to find two bins $L \in R_+$ and $H \in R_-$ such that $a_\ell \in L$ and $a_h \in H$. These bins $L$ and $H$ mark the attribute boundary bins. Then, $\mathcal{P}_{B|N,\mathcal{S}}(a_h)$ is identical to $R_+[H]$ and $\neg \mathcal{P}_{E|N,\mathcal{S}}(a)$ is identical to $R_-[L]$, where $R_+$ and $R_-$ are the bitmap indices described in Section \ref{sub:structure_and_of_the_hierarchical_bitmap_array_index}, each queried for a single bin. Then we add $\mathcal{P}_{N,\mathcal{S}}(a_\ell,a_h)$ to $p$ using bitwise OR.

Now, we process complete candidates in a similar fashion. Let $\mathcal{C}_{N,\mathcal{S}}(a_\ell,a_h)$ be a \emph{complete attribute match} bitmask specific to node $N$ for array of shape $\mathcal{S}$, so that the intersection $[a_\ell,a_h] \cap [min(N_i),max(N_i)] = [a_\ell,a_h]$.
\begin{align*}
\mathcal{C}_{N,\mathcal{S}}(a_\ell,a_h)[i] &= 1 \iff \mathcal{P}_{B|N,\mathcal{S}}(a_\ell)[i] \land \neg \mathcal{P}_{E|N,\mathcal{S}}(a_h)[i]
\end{align*}

This expression is very similar to $\mathcal{P}_{N,\mathcal{S}}(a_\ell,a_h)$, describing children that have started at or before $a_\ell$ and have not ended at or before $a_h$. To evaluate $\mathcal{C}_{N,\mathcal{S}}(a_\ell,a_h)$, we query $R_+[L]$ and $R_-[H]$. Then, we add the result to $c$ using bitwise OR and remove those from $p$, i.e., $p = p \land \neg c$.

Note that both partial and complete attribute candidates use a total of 4 index queries. An example of attribute query is displayed in the bottom row in Figure \ref{fig:partial-complete-query}.

\subsection{Dimension based Matches} % (fold)
\label{sub:dimension_based_matches}

Next, we explain how the dimension masks are set. This subsection further describes lines 9--17 in Algorithm \ref{algo_query}.

If for any dimension $d$ it holds that $d_h < D_{N_i}[d]_\ell$ or $a_\ell > D_{N_i}[d]_h$, there are neither partial nor complete dimension matches and we terminate processing the current node.

Unlike attribute query, the evaluation of dimension query is the same for all nodes $N$, so all the bitmaps for processing dimensions queries are \emph{precomputed}.

Let $\mathcal{P}'_{\mathcal{S},d}(d_\ell,d_h)$ be a \emph{partial dimension match}, where $d$ is a dimension in the query constraint $(d,d_\ell,d_h)$, for an array of shape $S$, indicating child nodes $N_i$ such that the intersection $[D_{N_i}[d]_\ell,D_{N_i}[d]_h,] \cap [d_\ell,d_h] \ne \emptyset$.

Let's fix a dimension $d$ for which we evaluate partial mat\-c0hes $\mathcal{P}'_{\mathcal{S},d}(d_\ell,d_h)$:
\begin{align*}
\mathcal{P}'_{\mathcal{S},d}(d_\ell)[i] &= 1 \iff d_\ell \in D_{N_i}[d] \land d_\ell \ne D_{N_i}[d]_\ell\\
\mathcal{P}'_{\mathcal{S},d}(d_h)[i] &= 1 \iff d_h \in D_{N_i}[d] \land d_h \ne D_{N_i}[d]_h \\
\mathcal{P}'_{\mathcal{S},d}(d_\ell,d_h)[i] &= 1 \iff \mathcal{P}'_{\mathcal{S},d}(d_\ell)[i] \lor \mathcal{P}'_{\mathcal{S},d}(d_h)[i]\\
\mathcal{P}'_{\mathcal{S}}(d_\ell,d_h)[i] &= \bigcup\limits_{1 \le d \le \texttt{DIMS}} \mathcal{P}'_{\mathcal{S},d}[i]
\end{align*}
The first expression describes which children $N_i$ have dimension $d$ range such that the query limit $d_\ell$ falls inside the range, but it is not equal to the lower limit of that range. The second expression is similar, but for $d_h$. Third and forth expression combine the partial matches over both query limits and all dimensions. Note that this results in excessive partial candidates since all child nodes that intersect the query constraints along at least one dimension qualify as partial candidates.

Partial dimension matches are evaluated using one precomputed bitmap index corresponding to
$$ \mathcal{P}'_{\mathcal{S},d}(b)[i] = 1 \iff b = D_{N_i}[d] $$
where $b$ is a bucket corresponding to the chunking of the array $\mathcal{A}$. There are a total of $F_d$ such buckets along dimension $d$, resulting in a total of $F_d \cdot d$ bitmaps of size $F$. We query these bitmaps for all dimensions and combine them using OR into $p'$

There is a special case of false negative dimension result. If $d_\ell$ or $d_h$ is equal to the d'th dimension range border of a child node $N_i$, and at the same time the other end of $d_\ell$ or $d_h$ causes the dimension to be fully covered in $N_i$, i.e. $d_\ell = D_{N_i}[d]_\ell$ and $d_h \ge D_{N_i}[d]_h$ or $d_h = D_{N_i}[d]_h$ and $d_\ell \le D_{N_i}[d]_\ell$, the query is evaluated as partial match for $N_i$ and dimension $d$, while in fact dimension $d$ contributes to complete matches. A check for this scenario requires comparing the dimension ranges of child nodes to the query range, and was ignored on purpose, as it complicates and slows down the query process.

For complete candidates, we will slightly modify the definition of $\mathcal{C}$ used for attributes. Let $\mathcal{C}'_{\mathcal{S},d}(d_\ell,d_h)$ be a \emph{complete dimension match} for array of shape $S$, indicating which child nodes $N_i$ are \emph{partially or fully} covered by interval $[d_\ell,d_h]$. Despite the semantics indicating partially matches should not be included, we later trim the complete dimension match bitmap accordingly.
\begin{align*}
\mathcal{C}'_{\mathcal{S},d}(d_\ell,d_h)[i] &= 1 \iff [d_\ell,d_h] \cap D_{N_i}[n] \ne \emptyset \\
\mathcal{C}'_{\mathcal{S}}(d_\ell,d_h)[i] &= \bigcap\limits_{1 \le n \le \texttt{DIMS}} \mathcal{C}'_{\mathcal{S},d}[i]
\end{align*}

Complete dimension matches are evaluated using two precomputed bitmap indices corresponding to
\begin{align*}
\mathcal{C}'_{B|\mathcal{S},d}(b)[i] &= 1 \iff b \le D_{N_i}[d]\\
\mathcal{C}'_{E|\mathcal{S},d}(b)[i] &= 1 \iff b \ge D_{N_i}[d]
\end{align*}
similarly to bitmaps used for partial matches. There is a total of $2 \cdot F_d \cdot d$ bitmaps of size $F$ for complete matches. We query these bitmaps for all dimensions and combine them using AND into $c'$.

\begin{figure}[htbp]
    \centering
    \includegraphics[width=0.48\textwidth]{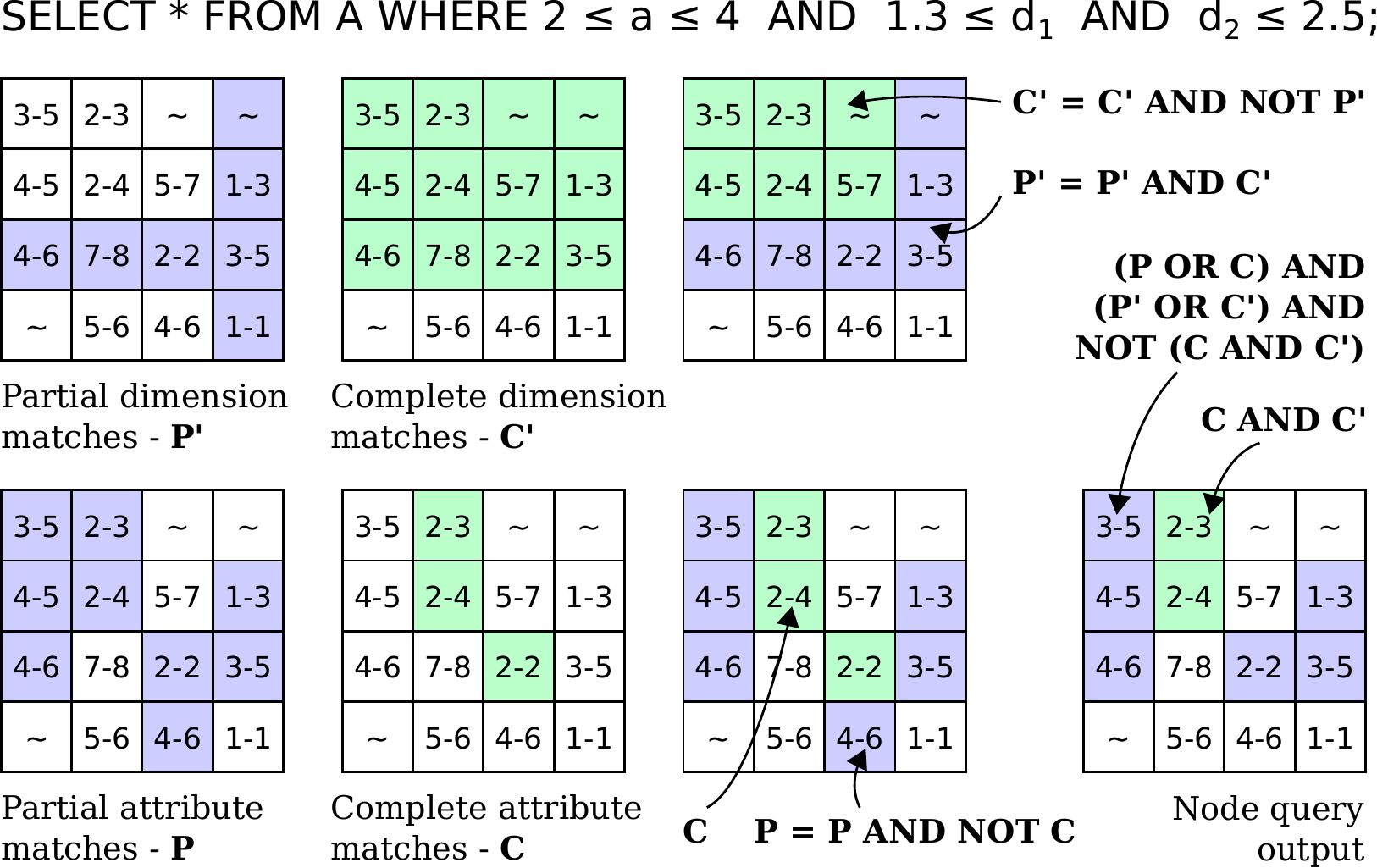}
    \vspace{1mm}
    \caption{Processing of a query in a single node of the hierarchical index. Top row represents dimension constraints, bottom row represents attribute constraints. Bottom right is the final product. Blue nodes represent partial matches and green node represent complete matches.}
    \label{fig:partial-complete-query}
\end{figure}

We now combine the partial dimension matches bitmap $c'$ with $p'$, such that $p' = p' \land c'$. Then, we clip the complete dimension bitmap by the partial bitmap as $c' = c' \land \neg p'$. During the evaluation of dimension matches, we used a total of $3 \cdot d$ index queries. An example of dimension query is displayed in the top row in Figure \ref{fig:partial-complete-query}.

\subsection{Partial and Complete Matches} % (fold)
\label{sub:partial_and_complete_matches}

Now that we have both attribute and dimension, and both partial and complete candidates, we may proceed to merging the candidates and generating a bitmap representing the set of result node children $\mathcal{C}_{N,\mathcal{S}}^*$ and a bitmap representing the set of potential node children $\mathcal{P}_{N,\mathcal{S}}^*$ that will be recursively explored. This subsection further describes lines 18--22 in Algorithm \ref{algo_query}.

The $\mathcal{C}_{N,\mathcal{S}}^*$ bitmap is easier to obtain, as it is the intersection of both complete bitmaps without partial candidates bitmaps.
$$\mathcal{C}_{N,\mathcal{S}}^* = \mathcal{C}_{N,\mathcal{S}} \land \mathcal{C}'_{\mathcal{S}}$$

We obtain the set of partial candidates $\mathcal{P}_{N,\mathcal{S}}^*$ by joining the dimension-based partial candidates with the attribute-based candidates and clipping both by complete candidates
$$\mathcal{P}_{N,\mathcal{S}}^* = (\mathcal{P}_{N,\mathcal{S}} \lor \mathcal{C}_{N,\mathcal{S}} ) \land (\mathcal{P}'_{\mathcal{S}} \lor \mathcal{C}'_{\mathcal{S}}) \land \neg \mathcal{C}_{N,\mathcal{S}}^*$$

We then iterate through the results, adding child nodes from $\mathcal{C}_{N,\mathcal{S}}^*$ to the result set $C$ and the partial candidates $\mathcal{P}_{N,\mathcal{S}}^*$ into the queue $P$ to be processed subsequently. This process is done on top of Z-order indices, as it is trivial to generate Z-order indices corresponding to nodes in the lower levels. The Z-order ordering of the inner nodes and breadth-first traversal also ensures single traversal through the index.

Running the algorithm for multiple queries or multiple attribute constraints in a single query can be implemented using iteration through the constraints in the worst case.

\subsection{Estimating Cardinality of Results; Membership Queries} % (fold)
\label{sub:estimating_min_max_count}

It is fairly straightforward to output estimates on minimal and maximal number of matching cells by iterating some bounded number of levels of the index. The minimal number outputs the size of nodes in $C$, while the maximum outputs the size of nodes in $C \cup P$. Using the $w(b)$ estimate, we may also provide estimates on aggregates over the attribute, based on bin-wise linear approximation.

There is a simple modification of the algorithm for membership queries. (See Section \ref{ssec:array_data_model} for details about membership queries). On top of two sided range indices $\mathcal{P}_{N,\mathcal{S}}$ and $\mathcal{C}_{N,\mathcal{S}}$ for attribute queries, we keep equality indices and iterate through the attribute constraint. For dimension membership queries, we precompute an index for all dimension values (within a single chunk), as opposed to buckets corresponding to child nodes, that are used in $\mathcal{P}'_{\mathcal{S},d}$ and $\mathcal{C}'_{\mathcal{S},d}$.

\section{Experimental Evaluation} % (fold)
\label{sec:experimental_evaluation}

We have tested our implementation against several other solutions, of which none is specifically tailored to mixed attribute and dimensions range queries, but those are the only readily available solutions involving bitmap indices and being capable of executing range queries.

We measured the time and space efficiencies for each individual query, i.e. total query execution time, and space requirements for the index. Timing was measured as an average of 3 runs with data preloaded into memory. For Fastbit queries, we use their internal wall time measuring systems, meaning certain pre and post processing steps are not included in the time measurements, such as query string parsing. Space requirements were measured based on the disk space required to store the bitmap index together with all relevant metadata.

The experiments were run on a single physical machine -- Intel(R) Xeon(R) CPU E5-1650 v2 @ 3.50GHz, 16 GB RAM, 1TB 7.2K RPM SATA 6Gbps; running Ubuntu 14.04.1 (3.19.0-32 kernel).

We use a synthetic dataset to test our queries on -- \emph{randomly generated multidimensional sum gaussian distribution} \textsc{SumGauss}. Its only attribute $a_G$ is a sum of $G$ randomly initialized Gaussian distribution in $D$ dimensions:
$$ a_G(\vec{d}) = \sum_{i=1}^{G}{\left(\frac{1}{\sqrt{(2\pi)^D|\Sigma_i|}}\exp{\left(-\frac{(d -\mu_i)^T\Sigma_i^{-1}(d-\mu_i)}{2}\right)}\right)}$$
where $\mu_i$ and $\Sigma_i$ are randomly generated distribution mean vector and a bounded symmetric positive definite covariance matrix for dimension $i$.
For sparse arrays, a threshold for the Gaussian functions is used. Attribute is treated as empty if the value is below this threshold. Only partitions with at least one non empty value are generated.

\subsection{Fastbit Integration} % (fold)
\label{sec:fastbit_integration}

Fastbit \cite{wu2009fastbit} is an open source library that implements bitmap indexing. It's not a complete database management system, rather a data processing tool, as its main purpose is to facilitate selection queries and estimates. Fastbit's key technological features are WAH bitmap compression multi-component and multi-level indices with many different combinations of encoding and binning schemes.

We use Fastbit's partitions to setup the lowest level of our indices (leaves), and base our binning indices on Fastbit's single-level binning index. This approach requires preprocessing of the data into evenly shaped partitions, generating empty bitmasks and shape metadata. Once a table is preprocessed into even partitions, it is indexed as described in Section \ref{sec:hierarchical_bitmap_array_index}. The index generation processes one partition at a time, and once processed, the partition is never accessed again during the index generation.

\subsection{Bitmap Indexing Methods} % (fold)
\label{sub:bitmap_indexing_methods}

\textsc{BoxClip} represents a naive algorithm using 32 equi-depth binned indices, interval encoding and WAH compression.
The result bitmask from the attribute query is transformed to a set of ``line'' hyperrectangles (size of the hyperrectangle in all but one dimensions is 1), which are filtered from the dimension query, then merged into a set of result hyperrectangles.
All the steps except filtering are built on top for Fastbit's mesh query. The filtering is implemented using recursive sweeping line algorithm.

\textsc{DimsAtts} uses indexed \texttt{uint} auxiliary attributes made from dimensions (see Section \ref{sec:hierarchical_bitmap_array_index}). The dimension query is preprocessed into attributes, then run as a multi constraint query in Fastbit. The configuration is the same as in \textsc{BoxClip}, using 32 binned indices, range encoding and WAH compression on all attributes.

\textsc{ArrayBit} represents our hierarchical multidimensional index. We use 16 equi-depth binned indices, range encoding and WAH compression to index the partitions, and 16 approximately equi-depth binned indices (described in Section \ref{sub:bin_boundaries_merging_in_the_parent_node}) with two sided range encoding and no compression for the hierarchical index. Note that compared to \textsc{BoxClip} and \textsc{DimsAtts}, we only use half of the bins in the partition index. It is sufficient in our algorithm, because the bin boundaries are adapted to the actual data in each partition, and because we need to store the bin boundaries within the partitions.

\subsection{Range Queries} % (fold)
\label{sub:range_queries}

In our work, we focus on mixed attribute and dimension queries. Regardless of the dataset, we categorize the queries based on the overall ratio of the size of the query result to the size of the total array size.

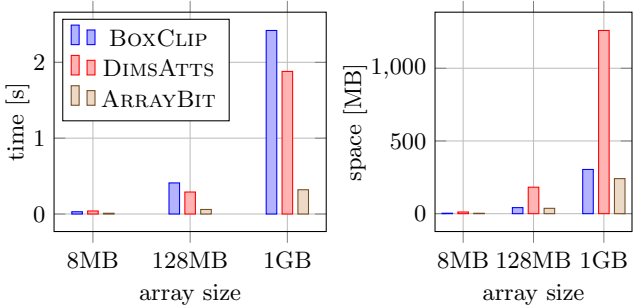
\begin{figure}[t!]
\begin{subfigure}[t]{0.24\textwidth}
    \centering
    \hspace{-3mm}\begin{tikzpicture}
    \begin{axis}[
        ybar,
        bar width = 4pt,
        height=4.5cm,
        width=5.2cm,
        grid=major,
        symbolic x coords={8MB,128MB,1GB},
        xtick=data,
        xlabel={array size},
        ylabel={time [s]},
        enlarge x limits=0.2,
        ylabel style={
            yshift = -5pt},
        yticklabel style={
            /pgf/number format/fixed,
            /pgf/number format/precision=1},
        legend pos=north west
    ]

    \addplot coordinates {
        (8MB,  0.03)
        (128MB, 0.41)
        (1GB, 2.42)
    };
    \addlegendentry{\textsc{BoxClip}}

    \addplot coordinates {
        (8MB,  0.04)
        (128MB, 0.29)
        (1GB, 1.88)
    };
    \addlegendentry{\textsc{DimsAtts}}

    \addplot coordinates {
        (8MB,  0.01)
        (128MB, 0.06)
        (1GB, 0.32)
    };
    \addlegendentry{\textsc{ArrayBit}}

    \end{axis}
    \end{tikzpicture}
    % \caption{Query execution time}
    \label{fig:time-queries}
\end{subfigure}\hspace{-1mm}%
\begin{subfigure}[t]{0.24\textwidth}
    \centering
    \begin{tikzpicture}
    \begin{axis}[
        ybar,
        bar width = 4pt,
        height=4.5cm,
        width=4.2cm,
        grid=major,
        symbolic x coords={8MB,128MB,1GB},
        xtick=data,
        xlabel={array size},
        enlarge x limits=0.2,
        ylabel={space [MB]},
        ylabel style={
            yshift = -5pt},
        yticklabel style={
            /pgf/number format/fixed,
            /pgf/number format/precision=1},
        legend pos=north west
    ]

    \addplot coordinates {
        (8MB,  3.01)
        (128MB, 41.56)
        (1GB, 304.10)
    };
    % \addlegendentry{\textsc{BoxClip}}

    \addplot coordinates {
        (8MB,  11.51)
        (128MB, 182.02)
        (1GB, 1260.05)
    };
    % \addlegendentry{\textsc{DimsAtts}}

    \addplot coordinates {
        (8MB,  2.83)
        (128MB, 36.80)
        (1GB, 241.05)
    };
    % \addlegendentry{\textsc{ArrayBit}}

    \end{axis}
    \end{tikzpicture}
    % \caption{Size of the index file}
    \label{fig:space-queries}
\end{subfigure}
\vspace{-2mm}
\caption{Query execution time and disk space required to store the indices for different array sizes.}
\label{fig:time-space-queries}
\vspace{2mm}
\end{figure}

Figure \ref{fig:time-space-queries} shows the time required to return all results. The index file is preloaded into memory prior to the test for all the systems used. We used 2D array for this experiment. and a query with $\approx 10\%$ hit ratio. Both \textsc{BoxClip} and \textsc{DimsAtts} run slower than \textsc{ArrayBit}. In case of \textsc{BoxClip}, the reason is that all the attribute query results had to be processed, while for \textsc{DimsAtts} the reason is that the attribute made from second dimension didn't effectively compress. In terms of space requirements, all of the algorithms save attribute index. \textsc{ArrayBit} uses less bins in the leaves, but stores bin boundaries for all leaves and internal nodes, plus bitmaps for internal nodes, effectively taking up the same space as \textsc{BoxClip}. On the other hand, \textsc{DimsAtts} stores indices for all dimension attributes. Row major ordering is used in this measurement.

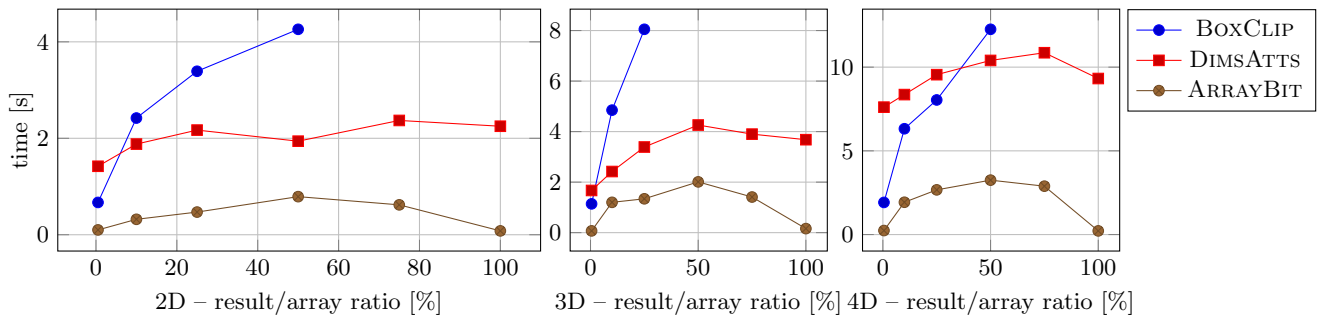
\begin{figure*}[t!]
    \centering
    \vspace{-2mm}
    \begin{tikzpicture}
    \begin{axis}[
        height=4.8cm,
        width=8cm,
        grid=major,
        xlabel={2D -- result/array ratio [\%]},
        ylabel={time [s]},
        ylabel style={
            yshift = -5pt},
        yticklabel style={
            /pgf/number format/fixed,
            /pgf/number format/precision=1}
    ]

    \addplot coordinates {
        (0.5,  0.67)
        (10.0, 2.42)
        (25.0, 3.39)
        (50.0, 4.26)
        % (75.0, 0.6)
        % (100.0,0.8)
    };
    % \addlegendentry{\textsc{1 KiB}}

    \addplot coordinates {
        (0.5,  1.42)
        (10.0, 1.88)
        (25.0, 2.17)
        (50.0, 1.94)
        (75.0, 2.37)
        (100.0,2.25)
    };
    % \addlegendentry{\textsc{32 KiB}}

    \addplot coordinates {
        (0.5,  0.10)
        (10.0, 0.32)
        (25.0, 0.47)
        (50.0, 0.79)
        (75.0, 0.62)
        (100.0,0.08)
    };
    % \addlegendentry{\textsc{1 MiB}}

    \end{axis}
    \end{tikzpicture}%
    \begin{tikzpicture}
    \begin{axis}[
        height=4.8cm,
        width=5cm,
        grid=major,
        xlabel={3D -- result/array ratio [\%]},
        ylabel style={
            yshift = -5pt},
        yticklabel style={
            /pgf/number format/fixed,
            /pgf/number format/precision=1}
    ]

    \addplot coordinates {
        (0.5,  1.14)
        (10.0, 4.85)
        (25.0, 8.05)
        % (50.0, 4.26)
        % (75.0, 0.6)
        % (100.0,0.8)
    };
    % \addlegendentry{\textsc{1 KiB}}

    \addplot coordinates {
        (0.5,  1.67)
        (10.0, 2.42)
        (25.0, 3.39)
        (50.0, 4.26)
        (75.0, 3.90)
        (100.0,3.68)
    };
    % \addlegendentry{\textsc{32 KiB}}

    \addplot coordinates {
        (0.5,  0.07)
        (10.0, 1.20)
        (25.0, 1.34)
        (50.0, 2.01)
        (75.0, 1.41)
        (100.0,0.16)
    };
    % \addlegendentry{\textsc{1 MiB}}

    \end{axis}
    \end{tikzpicture}\hspace{-5mm}
    \begin{tikzpicture}
    \begin{axis}[
        height=4.8cm,
        width=5cm,
        grid=major,
        xlabel={4D -- result/array ratio [\%]},
        ylabel style={
            yshift = -5pt},
        yticklabel style={
            /pgf/number format/fixed,
            /pgf/number format/precision=1},
        legend pos=outer north east
    ]

    \addplot coordinates {
        (0.5,  1.92)
        (10.0, 6.32)
        (25.0, 8.04)
        (50.0, 12.26)
        % (75.0, 8.04)
        % (100.0,0.8)
    };
    \addlegendentry{\textsc{BoxClip}}

    \addplot coordinates {
        (0.5,  7.61)
        (10.0, 8.36)
        (25.0, 9.55)
        (50.0, 10.40)
        (75.0, 10.86)
        (100.0,9.32)
    };
    \addlegendentry{\textsc{DimsAtts}}

    \addplot coordinates {
        (0.5,  0.24)
        (10.0, 1.93)
        (25.0, 2.67)
        (50.0, 3.25)
        (75.0, 2.89)
        (100.0,0.22)
    };
    \addlegendentry{\textsc{ArrayBit}}

    \end{axis}
    \end{tikzpicture}
\vspace{2mm}
\caption{Query execution time for 2D, 3D and 4D queries of various hit ratios. Queries contained an attribute constraint and all dimension constraints, each constraint with approximately the same domain reduction.}
\label{fig:time-based-on-dimensions}
\vspace{-2mm}
\end{figure*}

Figure \ref{fig:time-based-on-dimensions} demonstrates the dependency of the query processing time on a hit ratio of the query, i.e., the ratio of selected cells vs total cells in the array. \textsc{BoxClip} algorithm does not prune the search space based on the dimensions, resulting in number of hits dependent on the attribute only. Filtering these is is time intensive. \textsc{DimsAtts} depends linearly on the total number of dimensions. This is because there is an additional attribute for each dimension. There is also a small dependency on the hit ratio, where the increase is due to the results retrieval. \textsc{ArrayBit} achieves very good results for low or high hit rate queries. This is due to a large number of complete matches, and due to fast pruning of search space. For medium hit rate queries, the algorithm has relatively high number of candidate nodes to explore, but still manages to prune the search space faster.

\subsection{Parameterization} % (fold)
\label{sub:parameterization}

We also experimented with different setups of our hierarchical index. The major objectives remain the same: query execution time and space requirements of the index.

First, the \emph{partition size} determines the ratio of partition index vs hierarchical index. We set this in equilibrium with \emph{number of index bins}, which increases the precision of the binning and results in higher probability of pruning the search space earlier.

Another important parameter is a \emph{fanout} of nodes. If we use a smaller fanout (the smallest possible is $2^D$), we may not fill a single memory word with the index, significantly impair bit parallelism, furthermore the index size will be larger due to much deeper indexing tree. If the fanout is too high, we will not prune infeasible candidates fast enough. We got optimal results with a fanout close to a multiple of the word size, such as $8^2 = 64$ for 2D arrays, $4^3 = 64$ for 3D, $4^4 = 256$ for 4D, $3^5 = 243$ for 5D, etc.

\section{Conclusions and Future Work}
\label{sec:conclusion}

Most of the work on bitmap indexing to date focus on improving the space efficiency and speed, while a few applied the bitmap indices to multidimensional data. However, the linear form of bitmap indices was never adapted to support multidimensional array data.

We have proposed a bitmap indexing method that is designed for multidimensional arrays and focuses on overcoming the dimensionality issue. The hierarchical nature of the proposed method allows for continuous results and estimates to be output as intermediate results. Our approach effectively prunes the search space, uses data adaptive, approximate equi-depth binning. Furthermore, the index supports partitioned array data and allows distributed storage.

Our experimental results show that the proposed bitmap indexing method outperforms standard linearized approaches for mixed attribute and dimension range query processing.

There is a possible caveat that more complex multi-level and multi-component indices exist. None of these indices overcome the problem of dimensionality, rather due to their effectiveness delay the threshold where the drawbacks became noticeable (in terms of number of dimensions and size of the array).

Future work includes adapting the tree structure based on dimensions, such as adaptive mesh refinement widely used in physical simulations \cite{berger1989local}. Another interesting possibility is multi-attribute index in a single hierarchical structure. Last, we want to use better approximation algorithms to determine feasible regions from finer attribute bins.

\section{Acknowledgement}
This research was supported in part by AcRF Grant RG-18/14.

\vfill
\pagebreak
\bibliographystyle{abbrv}
\scriptsize
\bibliography{sigspatial2016,mendeley}
\end{document}